\documentclass[11pt,runningheads]{llncs}
\usepackage[left=1.5in,right=1.5in,top=1.5in,twoside=false]{geometry}
\usepackage{graphicx}
\usepackage{color}
\usepackage{caption,subcaption}
\usepackage[utf8x]{inputenc} 
\usepackage{amsmath} 
\usepackage{verbatim}
\usepackage[noabbrev]{cleveref}
\usepackage{url}
\usepackage{mathtools}
\usepackage{amssymb}
\usepackage{placeins}
\usepackage{algorithm, algorithmic}
\newtheorem{observation}{Observation}

\DeclarePairedDelimiter{\abs}{\lvert}{\rvert}

\DeclareMathOperator*{\prob}{\mathbb{P}}
\DeclareMathOperator*{\pop}{\mathcal{H}_{ab}}
\DeclareMathOperator*{\unif}{\mathcal{U}}

\allowdisplaybreaks

\begin{document}
\mainmatter  
\title{Construction and Random Generation of Hypergraphs with Prescribed Degree and Dimension Sequences}
\titlerunning{Construction and Random Generation of Hypergraphs}
\author{%
	Naheed Anjum Arafat\inst{1},
	Debabrota Basu\inst{2},\\
    Laurent Decreusefond\inst{3},
	St\'ephane Bressan\inst{1}
}%
\institute{
	School of Computing, National University of Singapore, Singapore
	\and
	Data Science and AI Division, Chalmers University of Technology, Sweden
	\and 
	LTCI, T\'el\'ecom Paris, Institut Polytechnique de Paris, France
}
\maketitle

\begin{abstract}
We propose algorithms for construction and random generation of hypergraphs without loops and with prescribed degree and dimension sequences. The objective is to provide a starting point for as well as an alternative to Markov chain Monte Carlo approaches. 
Our algorithms leverage the transposition of properties and algorithms devised for matrices constituted of zeros and ones with prescribed row- and column-sums to hypergraphs.  The construction algorithm extends the applicability of Markov chain Monte Carlo approaches when the initial hypergraph is not provided. The random generation algorithm allows the development of a self-normalised importance sampling estimator for hypergraph properties such as the average clustering coefficient. 

We prove the correctness of the proposed algorithms. We also prove that the random generation algorithm generates any hypergraph following the prescribed degree and dimension sequences with a non-zero probability. We empirically and comparatively evaluate the effectiveness and efficiency of the random generation algorithm. Experiments show that the random generation algorithm provides stable and accurate estimates of average clustering coefficient, and also demonstrates a better effective sample size in comparison with the Markov chain Monte Carlo approaches. 

\end{abstract}

\section{Introduction}
While graphs are the prevalent mathematical models for modern applications, being natural representations of varied objects such as transportation, communication, social and biological networks~\cite{newman}, to mention a few, they only capture binary relationships. Hypergraphs introduce the opportunity to represent $n$-ary relationships and thus create a more general, albeit more complex and generally more computationally expensive, alternative~\cite{berge1985graphs,klamt2009hypergraphs,wang}. 

Indeed many real-world systems are more naturally modelled as hypergraphs, as exemplified by the cases of multi-body systems, co-authorship networks and parliamentary relations~\cite{scholkopf,tan,wang,yang,benson2018simplicial}. While the applications are numerous, the properties of the underlying hypergraphs are yet to be fully understood. Just as it is the case for graphs in network science~\cite{fosdick2018configuring,blitzstein2011sequential,milo2003uniform}, configuration modelling or the random generation of hypergraphs with prescribed degree and dimension sequences precisely allows the fabrication of suitable hypergraphs for empirical and simulation-based studies of hypergraph properties~\cite{chodrow2019configuration}.

We study and propose algorithms for construction and random generation of hypergraphs with prescribed degree and dimension sequences (Sections~\ref{subsec:const},~\ref{subsec:rangen}). In addition, we present the necessary background on hypergraphs and $(0,1)$-matrices in Section~\ref{sec:back} and synthesise related works in Section~\ref{sec:related}.

Recently, Chodrow~\cite{chodrow2019configuration} proposed a Markov chain Monte Carlo (MCMC) algorithm to address this problem of generating a labelled hypergraph with a prescribed degree and dimension sequences. The MCMC algorithm requires an initial hypergraph with the prescribed degree and dimension sequences as a starting point. It is not always the case that such an initial hypergraph is available.
Therefore,  we present in Section~\ref{subsec:const} a deterministic algorithm for constructing an initial hypergraph as a starting point for the existing MCMC approach. At each iteration, our algorithm constructs the edge with the largest dimension using distinct vertices having the largest degrees. 

We present  in Section~\ref{subsec:rangen}  a random generation algorithm for generating hypergraphs as an alternative to the existing MCMC approach.
Our generation algorithm leverage properties and methods devised for $(0,1)$-matrices~\cite{ryser} with row- and column-sums coinciding with the hypergraph specification. If no row or column in the matrix contains only zeros, every $(0,1)$-matrix corresponds to the incidence matrix of a hypergraph with parallel-edges but no loop~\cite[Chapter~17]{berge1985graphs}. The column-sums of an incidence matrix represent degrees of the vertices and the row-sums represent dimensions of the edges of a hypergraph. 
At each iteration, the random generation algorithm generates the edge with the largest dimension using distinct, randomly selected vertices that satisfy the characterisation theorem for $(0,1)$ matrices (Theorem~\ref{thm:gale}). 

We further leverage our random generation algorithm to propose a self-normalised importance sampling (SNIS) estimator~\cite{kong1992note} for estimating hypergraph properties in Section~\ref{sec:sis}.

We prove the correctness of both the algorithms (Theorems~\ref{thm:cor_cons} and~\ref{thm:cor_gen}). Furthermore, we prove that the generation algorithm generates any random hypergraph having prescribed degree and dimension sequences with non-zero probability (Theorem~\ref{thm:prob_gen}).  
We evaluate the effectiveness (Section~\ref{sec:emp}) of the MCMC algorithm enabled with our construction algorithm and the random generation algorithm with SNIS estimator by estimating the average clustering coefficient of the projected graphs of the family of hypergraphs having prescribed degree and dimension sequence and also computing corresponding effective samples size~\cite{kong1992note}.

We conclude in Section~\ref{sec:conc} by summarising our findings.

\vspace*{-0.5em}
\section{Hypergraphs and $(0,1)$-matrices}
\label{sec:back}\vspace*{-0.5em}
In this section, we describe selected concepts of hypergraphs and inequalities involving $(0,1)$-matrices relevant for the transposition of properties and algorithms for $(0,1)$-matrices to hypergraphs. 


\begin{definition}[Hypergraph~\cite{berge1985graphs}]
A \textit{hypergraph} $H = (V, E)$ is a tuple of a vertex set $V = \{v_1,\ldots,v_n\}$ and an edge set $E = \{e_1,\ldots,e_m\}$ where each edge is a subsets of $V$.
\end{definition}
Unless otherwise stated, hypergraphs are \textit{labelled}, may contain \textit{parallel-edges}\footnote{$E$ is a multi-set}, but \textit{no self-loop}.
The polyadic relations i.e. the edges connecting the vertices in a hypergraph is presentable as a $(0,1)$-matrix, called the incidence matrix.
\begin{definition}[Incidence Matrix]
The incidence matrix $M = [m_{ij}]_{i,j=1,1}^{m,n}$ of a labelled hypergraph $H = (V, E)$ is a $(0,1)$-matrix with columns representing labels of vertices in $V$ and rows representing edge set $E$ where 
\[
m_{i,j} = \begin{cases}
1 \quad \text{ if } v_j \in e_i,\\
0 \quad \text{otherwise.} 
\end{cases}
\]
\end{definition}
The incidence matrix of a hypergraph is not unique. Even if the vertices are arranged in a total-order, such as in descending order of degrees and lexicographic order of labels among the vertices with same degree, any permutation of the rows would provide another incidence matrix of the same hypergraph. 

\noindent\textbf{Property 1.} Every incidence matrix of a hypergraph whose degree sequence is $(a)_n$ and dimension sequence is $(b)_m$ is contained in the set of $(0,1)$-matrices of dimension $m \times n$ whose column-sums are $(a)_n$ and row-sums are $(b)_m$. \textit{Thus, any algorithm that uses the characterisation of sequences $(a)_n$ and $(b)_m$ to construct an $m \times n$-dimensional $(0,1)$-matrix with column-sum $(a)_n$ and row-sums $(b)_m$ can be leveraged to construct a hypergraph with degree-sequence $(a)_n$ and dimension sequence $(b)_m$}. 
This observation constitute the core of our random hypergraph generation proposed in Section~\ref{subsec:rangen}. 



\noindent\textbf{Property 2.} In order to design the proposed algorithms and to prove their correctness, we would use the Gale-Rysers characterisation of $(0,1)$-matrices (Theorem~\ref{thm:gale}). Before discussing the theorem, we intend to remind us the notion of dominance between sequences.
\begin{definition}[Dominance~\cite{marshall1979inequalities}]\label{def:dom}
\emph{$(a)_n$ is defined to be dominated by $(b)_m$} if the sequences are zero-padded up to length $l=\max\{m,n\}$ to yield $(a^*)_l$ and $(b^*)_l$ that satisfy the following conditions: \\
\emph{(i)} sum of the first $k$ components of $(a^*)_l$ is smaller than or equal to sum of the first $k$ components of $(b^*)_l$ and\\
\emph{(ii)} sum of all the components of $(a^*)_l$ is equal to the sum of all the components of $(b^*)_l$. Mathematically,
\begin{align*}
(a)_n \prec (b)_m \Longleftrightarrow
\begin{dcases}
	\qquad\sum_{i=1}^k a^*_i &\leq \quad\sum_{i=1}^k b^*_i,\qquad k = 1,2,\ldots,l-1 \\
	\qquad\sum_{i=1}^{l} a^*_i &= \quad \sum_{i=1}^{l} b^*_i.\\
\end{dcases}    
\end{align*}
where $a^*_i = a_i$ for $i \leq n$, $a^*_i = 0$ for $i > n$, $b^*_i = b_i$ for $i\leq m$ and $b^*_i = 0$ for $i>m$.
\end{definition}

\begin{theorem}[Gale-Rysers~\cite{gale,ryser} Characterisation of Matrices]\label{thm:gale}
	If $(a)_n = (a_1,a_2,\ldots,$ $a_n)$ and $(b)_m = (b_1,b_2,\ldots,b_m)$ are two monotonically non-increasing, non-negative integer sequences, the necessary and sufficient condition for the existence of a $(0,1)$-matrix with column sums $(a)_n$ and row sums $(b)_m$ is that $(a)_n$ is dominated by the conjugate sequence of $(b)_m$.
\end{theorem}
The conjugate sequence of $(b)_m$ is a sequence whose $i^{th}$ component is the number of components in $(b)_m$ that are greater than or equal to $i$. We denote the conjugate sequence of $(b)_m$ as $\bar{b}_n$.\footnote{Since the number of 1's in a row of an $m \times n$-dimensional $(0,1)$ matrix cannot exceed $n$, the length of the conjugate sequence of row sums $(b)_m$ is upper bounded by $n$.}
A sequence-pair $((a)_n,(b)_m)$ satisfying the dominance condition in Gale-Rysers characterisation is said to be \textit{realisable by a $(0,1)$-matrix.} Conversely, such a matrix is said to \textit{realise} $(a)_n,(b)_m$.

\noindent\textbf{Property 3.} Another observation is that if we construct a new sequence $(a')_{n-1} = (a_2,a_3,\ldots,a_n)$  from a monontonically non-increasing positive integer sequence $(a)_n = (a_1,a_2,a_3,\ldots,a_n)$, the conjugate sequence of $(a)_{n-1}$ can be derived from the conjugate sequence of $(a)_n$ by reducing the first $a_1$ components of $\bar{a}$ by $1$.
\begin{lemma}[\cite{marshall1979inequalities}]
	\label{lemma:maj}
	Let $(a)_n = (a_1,a_2,\ldots,a_n)$ be a positive monotonically non-increasing. If we construct a new sequence $(a')_{n-1} \triangleq (a_2,\ldots, a_n)$, then the conjugate sequence of $(a')_{n-1}$ is
	$$ (\bar{a'}) = (\bar{a}_1 - 1,\ldots, \bar{a}_{a_1}-1,\bar{a}_{a_1+1}, \ldots,\bar{a}_n).$$
\end{lemma} 
\begin{example}
	Let $(a) = (4,2,2,1)$. Its conjugate sequence is $(\bar{a}) = (4,3,1,1,0,\ldots)$. By removing $a_1$, we get a new sequence $(a') = (2,2,1)$. The conjugate sequence of $(a')$ is $(3,2,0,0,\ldots)$ which is exactly the sequence derived from $(4,3,1,1,0,\ldots)$ by reducing first four components by 1 i.e. $(4-1,3-1,1-1,1-1,0,\ldots)$.
\end{example}
Fulkerson and Ryser~\cite{fulkersonryser} state a necessary condition that preserves dominace after reducing the values of a fixed number of components by 1 in sequences $(a)_n$ and $(b)_n$ related by dominance.
\begin{lemma}[Fulkerson-Ryser's Lemma~\cite{fulkersonryser}]
    \label{lemma:gfr}
	Let $(a)_n$ and $(b)_n$ be two monotonically non-increasing integer sequences. Let $(u)_n$ be sequence obtained from $(a)_n$ by reducing components at indices $i_1,i_2,\ldots,i_K$ by 1. Similarly, let $(v)_n$ be obtained from $(b)$ by reducing components at indices $j_1,j_2,\ldots,j_K$ by 1.
	
	If $i_1 \leq j_1, i_2 \leq j_2, \ldots, i_K \leq j_K$, and  $(a)_n \prec (b)_n$, we get $(u)_n \prec (v)_n$.
\end{lemma}
We leverage this lemma to prove correctness of our construction algorithm (Section~\ref{subsec:rangen}).

\section{Related Works}\label{sec:related}
\subsection{Graphs with a Prescribed Degree Sequence}
There are two main frameworks for the generation of random graphs with a prescribed degree sequence. The first framework is \textit{direct sampling}~\cite{fosdick2018configuring}, that constructs the graph incrementally edge-by-edge. 
Among algorithms based on direct sampling, \cite{bollobas1980probabilistic} and~\cite{bender1978asymptotic} introduced the concept of \textit{stubs} and the procedure of stub-matching as an algorithm for counting the number of labelled graphs with a prescribed degree sequence. The stub-matching procedure may generate graphs with loops and parallel-edges, which is often undesirable. Rejecting the generated random graph until a simple graph is generated is proposed as a remedy. However, this approach is inefficient for large degree values as an exponential number of samples might get rejected~\cite{blitzstein2011sequential,fosdick2018configuring}. Furthermore, there is no obvious way to extend this algorithm for graphs into an algorithm for hypergraphs~\cite{chodrow2019configuration}.

As an alternative to the stub-matching algorithm, \cite{blitzstein2011sequential} proposed an algorithm that uses the Erd\"{o}s-Gallai's characterisation to generate simple graphs. This algorithm generates all simple graphs following a given degree sequence with a non-zero probability. \cite{blitzstein2011sequential} also proposes an importance sampling scheme to estimate the number of simple graphs following the prescribed degree sequence. Motivated by their work on simple graphs, in this paper, we devise a self-normalised importance sampling scheme (Section~\ref{sec:sis}) using our random generation algorithm (Section~\ref{subsec:rangen}) to estimate average clustering coefficient of projected graphs of hypergraphs having a prescribed degree and dimension sequences (Section~\ref{sec:emp}). 

The second framework proposes \textit{MCMC} algorithms~\cite{milo2003uniform,newman,fosdick2018configuring} that iteratively switch edges of an initial graph with a given degree sequence to obtain the final graph. MCMC algorithms try and show that the intermediate hypergraphs form a Markov chain whose stationary distribution converges to the uniform distribution over the set of all graphs with the given degree sequence~\cite{fosdick2018configuring}. However, it is challenging to prove mixing-time bounds for algorithms in this family, and mixing results are known only for a limited case of regular graphs~\cite{blitzstein2011sequential}. We discuss an echo of this issue for hypergraphs in Section~\ref{sec:sis}.
\subsection{Hypergraphs with Prescribed Degree and Dimension Sequences}
Chodrow~\cite{chodrow2019configuration} proposed a hypergraph configuration modelling approach to the uniform distribution of labelled hypergraphs with prescribed degree and dimension sequence. The hypergraphs under investigation have parallel-edges but no self-loop. He proposed an MCMC algorithm that, as it is done in similar algorithms for graphs, sequentially switches edges of a labelled initial hypergraph satisfying the prescribed degree and dimension sequences. As the lag at which to sample a hypergraph from the Markov chain tends to infinity, they show that the algorithm outputs uniformly at random a hypergraph from the set of all hypergraphs having the prescribed degree and dimension sequences.

However, in practice, the initial hypergraph is not always available. Additionally, due to lack of mixing time results about the chain, there is no principled guideline for the lag at which a practitioner would sample a hypergraph from the chain. These observations motivated us to develop both a deterministic algorithm to construct an initial hypergraph facilitating the MCMC algorithm, as well as a random generation algorithm that does not need an initial hypergraph as an alternative to the MCMC algorithm. 



\section{Construction of An Initial Hypergraph}\label{subsec:const}
We leverage the properties elaborated in Section~\ref{sec:back} to construct a hypergraph with prescribed degree and dimension sequences. This hypergraph provides a starting point for an MCMC algorithm~\cite{chodrow2019configuration}. Our algorithm uses the methodology proposed by Ryser~\cite{ryser} for $(0,1)$-matrices and by Gale~\cite{gale} for flows in bipartite-graphs. 
We illustrate the pseudocode in Algorithm~\ref{alg:constr}.
\textit{At each iteration, Algorithm~\ref{alg:constr} constructs the edge with the largest dimension using distinct vertices having the largest degrees.}





In Algorithm~\ref{alg:constr}, the aim is to construct a hypergraph with $n$ vertices, $m$ edges, degree sequence $(a)_n$, and dimension sequence $(b)_m$. Algorithm~\ref{alg:constr} takes non-increasingly sorted sequences $(a)_n$ and $(b)_m$ as input. It initialises $(a)^1$ as $(a)_n$ and $(b)^1$ as $(b)_m$. At each iteration $j \in \{1,\ldots,m\}$, it constructs an edge by selecting $b^j_1$ distinct vertices with maximal non-zero degrees\footnote{Here the ties are broken using the lexicographic order of the vertex-labels}. Then it constructs $(a)^{j+1}$ by reducing the degrees of the selected vertices in $(a)^j$ by $1$ and refers to $(b^j_2,\ldots, b^j_m)$ as $(b)^{j+1}$. It proceeds to construct the next edge using $(a)^{j+1}$ and $(b)^{j+1}$, and continues until all $m$ edges are constructed. 
 \begin{algorithm}[t!]
	\begin{algorithmic}[1]
		\REQUIRE{Degree and dimension sequences, $(a)_n$  and $(b)_m$, sorted in descending order}
		\ENSURE{Hypergraph $H = (V,E)$}
		\STATE \textbf{Initialise:} $V \leftarrow \{1,\ldots,n\}$, $E \leftarrow \phi$, $(a)^1 \leftarrow (a)_n$, $(b)^1 \leftarrow (b)_m$
		\FOR {$j = 1,2,\cdots, m$}
		\STATE Construct edge $e_j = \{v_1,\cdots, v_{b^j_1}\}$
		\STATE Construct $(a)^{j+1}_n$ by reducing the first $b^j_1$ components of $(a)^j_n$ by 1.
		\STATE Construct $(b)^{j+1} = (b_2^j,b_3^j,\ldots,b_m^j)$
		\STATE $E \leftarrow E \cup e_j$
		\STATE Sort sequence $(a)^{j+1}$ in descending order.
		\ENDFOR 
	\end{algorithmic}
	\caption{Constructing initial hypergraphs}\label{alg:constr}
\end{algorithm}

We prove that the construction of edge $e_j$ at every iteration $j$ is feasible, meaning, the residual sequences $(a)^{j+1}$ and $(b)^{j+1}$ after reduction are realisable by some hypergraph. 

\begin{theorem}
\label{thm:cor_cons}
If the sequences $(a)^j$ and $(b)^j$ are realisable by a hypergraph with $m$ edges and $n$ vertices, the sequences $(a)^{j+1}$ and $(b)^{j+1}$, constructed at iteration $j$, are realisable by a hypergraph with $(m-1)$ edges and $n$ vertices. 
\end{theorem} 
\noindent \textbf{Proof sketch.}
We prove the theorem by induction on $m$.

If $m = 0$, the algorithm terminates with a hypergraph with empty edges ($E=\phi$), which is the only hypergraph with $0$ edges and $n$ vertices. 

Suppose $m>0$. By induction hypothesis, $(a)^j,(b)^j$ are realisable by a hypergraph $H$ with $m$ edges. Taking an incidence matrix $M$ of $H$ and applying ~\Cref{thm:gale}, we get $(a)^j \prec (\bar{b})^j$. By construction, $(a)^{j+1}$ is the same as $(a)^j$ except the first $b^j_1$ components are reduced by 1. By construction of $(b)^{j+1}$ and Lemma~\ref{lemma:maj}, the conjugate $(\bar{b})^{j+1}$ is the same as $(\bar{b})^j$ except the first $b^j_1$ components reduced by 1. Thus Lemma~\ref{lemma:gfr} implies that $(a)^{j+1} \prec (\bar{b})^{j+1}$. By \Cref{thm:gale}, an $(m-1)\times n$-dimensional incidence matrix $M'$ of some hypergraph $H'$ exists that realises sequences $(a)^{j+1},(b)^{j+1}$.

The detailed proof is in Appendix~\ref{app:a}.

\vspace*{-1em}
\section{Random Generation of Hypergraphs}
\label{subsec:rangen}\vspace*{-.5em}
In this section, we propose a random generation algorithm (Algorithm~\ref{alg:gen}) using the characterisation (\Cref{thm:gale}) for $(0,1)$-matrices. In Algorithm~\ref{alg:gen}, we iteratively construct edges in descending order of cardinality and stochastically assign the vertices to the edges  such that Theorem~\ref{thm:gale} is satisfied. Algorithm~\ref{alg:gen} leverages design methods proposed for $(0,1)$-matrices in~\cite{chen2005sequential}.\\
Three observations are central to the development of Algorithm~\ref{alg:gen}.
\begin{observation}\label{obs:1}
If there are two sequences $(b)^j = (b^j_1,b^j_2,\ldots,b^j_m)$ and $(b)^{j+1} = (b^j_2,\ldots,b^j_m)$, Lemma~\ref{lemma:maj} implies that we can construct the conjugate sequence of $(b)^{j+1}$, namely $(\bar{b})^{j+1}$, from the conjugate sequence of $(b)^j$, namely $(\bar{b})^j$, by reducing first $b^j_1$ components of $(\bar{b})^j$ by 1. 
\end{observation}
\begin{observation}\label{obs:2}
If we randomly select $K$ non-zero components from $(a)^j$ whose indices are $i_1,\ldots,i_K$ and reduce them by $1$, we obtain a residual sequence $(a)^{j+1}$. \textit{If we select those $K$ components in such a way that after reduction the dominance $(a)^{j+1} \prec (\bar{b})^{j+1}$ holds, we can construct an $(m-1) \times n$-dimensional $(0,1)$-matrix with residual  column-sums $(a)^{j+1}$ and row-sums $(b)^{j+1}$.} This is direct consequence of Gale-Rysers theorem (\Cref{thm:gale}). The constructed $(0,1)$-matrix is an incidence matrix of a hypergraph with $m-1$ edges and $n$ vertices having degree sequence $(a)^{j+1}$ and dimension sequence $(b)^{j+1}$.
\end{observation}
\begin{observation}\label{obs:3}
Since our interest is to reduce $K$ non-zero components of $(a)^j$ by $1$ while preserving the dominance $(a)^{j+1} \prec (\bar{b})^{j+1}$, we search for the indices in $(a)^j$ where the violation of dominance $(a)^j \nprec (\bar{b})^{j+1}$ occur. 
We say an index $1 \leq k < n$ is \textbf{critical} if 
\vspace*{-.5em}
\begin{equation}\label{eqn:critical}
	    \sum_{i=1}^{k} a^j_i > \sum_{i=1}^{k} \bar{b}^{j+1}_{i}.\vspace*{-.5em}
\end{equation}
$k$ being a critical index implies that in order to preserve dominance $(a)^{j+1} \prec (\bar{b}^{j+1})$ within integer interval $[1,k]$, we need to reduce at least \vspace*{-.5em}
\begin{equation}\label{eqn:margin}
	    n \triangleq \sum_{i=1}^{k} a^j_i - \sum_{i=1}^{k} \bar{b}^{j+1}_{i}
\end{equation}
number of $1$'s at or before index $k$ in $(a)^j$. We say $n$ is the \textbf{margin of violation} corresponding to the critical index $k$. At every iteration, we enlist all the critical indices and their corresponding margins of violation. 
\end{observation}
	

 \begin{algorithm}[t!]
	\begin{algorithmic}[1]
		\REQUIRE{Degree and dimension sequences, $(a)_n$  and $(b)_m$, sorted in descending order}
		\ENSURE{Hypergraph $H = (V,E)$}
		\STATE \textbf{Initialise:} $V \leftarrow \{1,\ldots,n\}$, $E \leftarrow \phi$, $(a)^1 \leftarrow (a)_n$, $(b)^1 \leftarrow (b)_m$.
		\STATE $(\bar{b})^1 \leftarrow$ conjugate sequence of $(b)^1$
		\FOR {$j = 1,2,\cdots, m$}
		\STATE Construct $(\bar{b})^{j+1}$ from $(\bar{b})^j$ by reducing first $b^j_1$ components in $(\bar{b})^j$ by 1.
		\STATE Compute critical indices $\{k^j_1, k^j_2, \ldots\}$ where $(a)^j \nprec (\bar{b})^{j+1}$ (\Cref{eqn:critical}).
		\STATE Compute corresponding margins of violation $\{n^j_1, n^j_2, \ldots\}$ (\Cref{eqn:margin}).
		\STATE $e_j \leftarrow \phi$, $k^j_0 \leftarrow 0$
		\WHILE{$k^j_i \in \{k^j_1, k^j_2, \ldots\}$}
		    \STATE $o_i \rightarrow$ An integer sampled from $[n^j_{i}-|e_j|,\min(b^j_1 - |e_j|,k^j_{i}-k^j_{i-1})]$ uniformly at random.
		    \STATE $O_i \rightarrow o_i$ indices selected from $\mathcal{I}^j_i = [k^j_{i-1}+1,k^j_i]$ uniformly at random. 
		    \STATE $e_j \leftarrow e_j \cup O_i$ 
		    \STATE Reduce components in $(a)^j$ at positions $O$ by 1.
		\ENDWHILE
		\STATE $E \leftarrow E \cup e_j$.
		\STATE $(a)^{j+1} \leftarrow$ $(a)^j$ sorted in descending order.
		\STATE Construct $(b)^{j+1} = (b_2^j,b_3^j,\ldots,b_m^j)$.
		\ENDFOR 
	\end{algorithmic}
	\caption{\label{alg:gen} Generating random hypergraphs}
\end{algorithm}
Algorithm~\ref{alg:gen} takes the degree and dimension sequences,  $(a)_n$ and $(b)_m$ respectively, sorted in descending order as input. We refer to them as $(a)^1 = (a)_n$ and $(b)^1 = (b)_m$ (Line 1). 
Following that, it constructs the conjugate $(\bar{b})^1$ of the initial dimension sequence $(b)^1$ (Line 2). 

At each iteration $j \in \lbrace1, \ldots, m\rbrace$, the algorithm constructs a conjugate sequence for dimensions of $(m-j)$ edges, namely $(\bar{b})^{j+1}$, from the conjugate sequence for dimensions of $(m-j+1)$ edges, namely $\bar{b}^j$, by reducing the first $b_1^j$  components in $(\bar{b})^j$ by 1 (Line 4). This is a consequence of Observation~\ref{obs:1}.



Following Observation~\ref{obs:3}, Algorithm~\ref{alg:gen} uses $(a)^j$ and $(\bar{b})^{j+1}$ to compute all the critical indices $\{k^j_1,k^j_2,\ldots\}$ (Line 5) and their corresponding margins of violations $\{n^j_1,n^j_2,\ldots\}$ (Line 6). The critical indices partition $\{1, \ldots, n\}$ into integer intervals $\mathcal{I}^j_i \triangleq [k^j_{i-1}+1, k^j_{i}]$.

Now, we select indices from these partitions and aggregate them to resolve the critical indices. These selected indices construct a new edge $e_j$.
Following Observation~\ref{obs:2}, constructing edge $e_j$ reduces the problem of generating $(m-j+1)$ edges satisfying $(a)^{j}$ and $(b)^{j}$ to generating $(m-j)$ edges satisfying $(a)^{j+1}$ and $(b)^{j+1}$ conditioned on $e_j$. 

Specifically, in Line 7, the algorithm begins the edge construction considering the edge $e_j$ to be empty. In Lines 8-13, Algorithm~\ref{alg:gen} selects \textit{batches of vertices from integer interval $\mathcal{I}^j_i$ of indices and reduce $1$ from them till all the critical vertices $k^j_i$'s are considered}.
As these batches of vertices are selected, they are incrementally added to $e_j$.

Now, we elaborate selection of the batches of vertices from these intervals as executed in Lines 9-10. 
At the $i^{th}$ step of selecting vertices, the algorithm uniformly at random select $o_i$ indices from $\mathcal{I}^j_i$. $o_i$ is an integer uniformly sampled from the following lower and upper bounds: 
\begin{itemize}
\item \textit{Lower bound:} Since at least $n^j_i$ vertices have to be selected from $[1,k^j_i]$ to reinstate dominance and $|e_j|$ vertices have already been selected from $[1,k^j_{i-1}]$, the algorithm needs to select at least $n^j_i-|e_j|$ vertices from $\mathcal{I}^j_i$
\item \textit{Upper bound:} There are $(k^j_{i} -k^j_{i-1})$ indices in interval $\mathcal{I}^j_i$. After selecting $|e_j|$ vertices, the algorithm can not select more than $b^j_1 - |e_j|$ vertices. Thus, the maximum number of vertices selected from $\mathcal{I}^j_i$ is $\min(k^j_{i} -k^j_{i-1}, b^j_1 - |e_j|)$. 
\end{itemize}
Subsequently, the algorithm adds the $o_i$ vertices at those indices to the partially constructed $e_j$ (Line 11) and reduce the components at those selected indices in sequence $(a)^j$ by 1 (Line 12). 

After adding the edge $e_j$ to the edge set $E$ (Line 14), the algorithm sorts $(a)^{j}$ in descending order to construct $(a)^{j+1}$, removes $b_1^j$ from $(b)^j$ to construct $(b)^{j+1}$  (Line 15-16). In next iteration, the algorithm focuses on generating $(m-j)$ edges satisfying $(a)^{j+1}$ and $(b)^{j+1}$ conditioned on $e_j$.


In order to prove correctness of Algorithm~\ref{alg:gen}, we prove Theorems~\ref{thm:cor_gen} and~\ref{thm:prob_gen}.

\begin{theorem}
\label{thm:cor_gen}
If the sequences $(a)^j$ and $(b)^j$ are realisable by a hypergraph with $m$ edges and $n$ vertices, the sequences $(a)^{j+1}$ and $(b)^{j+1}$ as constructed by the algorithm at iteration $j$ are realisable by a hypergraph with $(m-1)$ edges and $n$ vertices. 
\end{theorem} 
\noindent\textbf{Proof sketch.}
This proof is similar to the proof of \Cref{thm:cor_cons} in spirit. The only difference is in the inductive step, where we need to prove that the choice of batches of vertices leading to sequences $(a)^{j+1}$ and $(b)^{j+1}$ is such that $(a)^{j+1} \prec (\bar{b})^{j+1}$. 

After reducing 1 from the selected indices in $(a)^j$, the resulting sequence $(a)^{j+1}$  must follow the inequality $\sum_{i=1}^k a_i^{j+1} \leq \sum_{i=1}^k \bar{b}_i^{j+1}$ at every index $k \in [1,n-1]$. Following Equation~\ref{eqn:margin}, if index $k$ is critical, $\sum_{i=1}^k a_i^{j+1} \leq \sum_{i=1}^k a_i^{j} - n_k = \sum_{i=1}^k \bar{b}^{j+1}$. If $k$ is not critical, $\sum_{i=1}^k a_i^{j+1} < \sum_{i=1}^k a_i^{j} \leq \sum_{i=1}^k \bar{b}^{j+1}$ by definition~\ref{eqn:critical}. After all the critical indices are considered, $\sum_{i=1}^n a_i^{j+1} = (\sum_{i=1}^n a_i^j) - b^j_1 = (\sum_{i=1}^n b_i^j) - b^j_1 = \sum_{i=1}^n b_i^{j+1}$. Consequently, we get that $(a)^{j+1} \prec (\bar{b})^{j+1}$.

\begin{theorem}
\label{thm:prob_gen}
Algorithm~\ref{alg:gen} constructs every hypergraph realisation of $(a)_n,(b)_m$ with a non-zero probability.
\end{theorem}
\noindent\textbf{Proof sketch.}
Let us begin with an arbitrary hypergraph realisation $H_1 = (V,E_1 = \{e_1,\ldots,e_m\})$ of sequences $(a)^1 = (a)_n,(b)^1 = (b)_m$ such that $|e_1| \geq |e_2| \geq \ldots \geq |e_m|$. 
At iteration 1, Algorithm~\ref{alg:gen} allocates vertices to edge $e_1$ with a probability 
$$\prob(e_1) = \frac{o_1}{k^1_1(\min(b^1_1,k^1_1) - n^1_1 + 1)}\frac{o_2}{(k^1_2-k^1_1)(\min(k^1_2-k^1_1,b^1_1-o_1) - (n^1_2-o_1) + 1)}\cdots$$ 

$\prob(e_1)$ is non-zero. Compute the conditional probabilities $\prob(e_2|e_1),\prob(e_3|e_2,e_1),\ldots,\linebreak \prob(e_m|e_{m-1},\ldots,e_1)$ in a similar manner. Each of the probabilities is non-zero, since \Cref{thm:cor_gen} implies that each of the intermediate sequence-pairs $((a)^j,(b)^j)$ is realisable. 
\textit{The joint probability with which the algorithm constructs the edge-sequence $E_1 \triangleq \lbrace e_1, \ldots, e_m \rbrace$} is 
$$\prob(E_1) \triangleq \prob(e_1) \prob(e_2|e_1) \ldots  \prob(e_m|e_{m-1},\ldots,e_1).$$ 
$\prob(E_1)$ being a product of non-zero terms is non-zero.

There are $c(E_1) = \frac{m!}{\prod_j mult^{E_1}(e_j)!}$ permutations of $e_1,\ldots,e_m$ that result in the same hypergraph as $H_1$. Here, $mult^{E_1}(e_j)$ is the multiplicity of edge $e_j$ in multi-set $E_1$. Let us denote the set of all permutations of $E_1$ by $[E_1]$. Thus, the algorithm constructs $H_1$ with probability $\prob(H_1) \triangleq \sum_{E\in [E_1]} \prob(E)$. $\prob(H_1)$ being a sum of non-zero terms, is non-zero. 

The detailed proofs are in Appendices~\ref{app:b} and~\ref{app:c}.

\section{Self-Normalised Importance Sampling Estimator}
\label{sec:sis}
In practice, it is desirable to apply a generation algorithm that samples hypergraphs from an uniform distribution over the population of hypergraphs $\pop$ having the prescribed degree and dimension sequences $(a)_n$ and $(b)_m$. Uniform generation is a desired property, as uniformly generated sample hypergraphs from $\pop$ can be used to estimate properties of the hypergraph population $\pop$. Also, the properties estimated from uniform distribution over population can be used to estimate properties for other distributions over the population. 

Yet enumerating all hypergraphs from the population $\pop$ is computationally infeasible as the problem of explicit enumeration of $(0,1)$-matrices with given row- and column-sums is \textbf{\#P}-hard~\cite{dyer}.
This result not only makes unbiased estimation of properties of $\pop$ computationally infeasible but also hardens the validation of uniformity or unbiasedness of any random generation algorithm.
Testing whether a random generation algorithm is uniform using state-of-the art algorithms for uniformity testing~\cite{batu} for the unknown discrete space $\pop$ is also computationally infeasible due to the astronomically large number of samples required by the testing algorithm.

The inaccessibility of population space $\pop$ motivates us to design an \textit{importance sampling based estimator} (SNIS). We use SNIS to estimate properties of hypergraphs in $\pop$ even if the induced distribution of generation algorithm is not uniform. Importance sampling~\cite{blitzstein2011sequential} assigns weights to estimates derived from non-uniformly generated hypergraph samples using the probability at which the hypergraphs are generated.

Let the uniform distribution over hypergraphs $H \in \pop$ be $\unif(H) \triangleq \frac{1}{\abs{\pop}}$. We are interested in estimating expected value $\mathbb{E}_{\unif}[f]$ of some hypergraph property $f: \pop \rightarrow \mathbb{R}$. For example, $f$ can be the average clustering coefficient of the projection of the hypergraphs to graphs. 
If we were able to access $\unif$ and draw $N$ i.i.d samples $H'_1,\ldots,H'_N$, the Monte Carlo estimate of $\mu(f) \triangleq \mathbb{E}_{\unif}[f]$ is $\frac{1}{N}\sum_{i=1}^N f(H'_i)$.  In practice, it is not feasible. 

Thus, we draw $N$ independent edge-sequences $E_1,\ldots,E_N$ from the space of edge-sequences $\mathcal{E}_{ab}$ leading to the hypergraphs in $\mathcal{H}_{ab}$. Using Algorithm~\ref{alg:gen}, we generate $N$ such edge-sequences $\lbrace E_i\rbrace_{i=1}^N$ with probabilities $\lbrace \prob(E_i)\rbrace_{i=1}^N$ respectively. We denote the hypergraph constructed by an edge-sequence $E_i$ as $H(E_i)$. We also observe that the uniform distribution $\unif$ over the space of hypergraphs $\mathcal{H}_{ab}$ induces a distribution $\hat{\unif}$ over the edge-sequences in $\mathcal{E}_{ab}$.


Following that, we evaluate property $f$ on the generated hypergraphs $\lbrace H(E_i) \rbrace_{i=1}^N$ and apply Equation~\ref{eq:is} to estimate the population mean $\mu(f)$.


\begin{equation}
\label{eq:is}
\hat{\mu}(f) = \frac{1}{N} \sum_{i=1}^{N} \frac{\hat{\unif}(E_i)}{\prob(E_i)} f(H (E_i)) \triangleq \sum_{i=1}^{N}w_i f(H(E_i))
\end{equation}
This is analogous to endowing an \textit{importance weight} $w_i$ to a sample $H(E_i)$. 
The sample mean $\hat{\mu}$ of population mean $\mu$ is unbiased. We provide a detailed discussion and a proof of this in Appendix~\ref{app:d}.

Computing $\hat{\mu}$ requires distribution $\hat{\unif}$ to be known. Since the cardinality of $\pop$ and thus that of $\mathcal{E}_{ab}$ are computationally unavailable, we adopt a self-normalised importance sampling estimator (SNIS) that uses normalised weights $w^{\mathrm{SNIS}}_i \triangleq \frac{w_i}{\sum_i w_i}$. Although SNIS is a biased estimator, it has been shown to work well in practice~\cite{blitzstein2011sequential,chen2005sequential}. We adopt SNIS for statistical property estimation~\cite{kong1992note}, and define SNIS estimator $\tilde{\mu}$ for a hypergraph property $f$ as 
\begin{align}\label{eqn:SNIS}\hspace*{-1.2em}
    \tilde{\mu}(f) &\triangleq \frac{\sum_{i=1}^N \frac{\hat{\unif}(E_i)}{\prob(E_i)} f(H(E_i))}{\sum_{i=1}^{N} \frac{\hat{\unif}(E_i)}{\prob(E_i)}} \notag\\
    &= \sum_{i=1}^N \frac{1}{\prob(E_i)(\sum_{i=1}^N \frac{1}{\prob(E_i)})}f(H(E_i)) \notag \\
    &\triangleq \sum_{i=1}^N w^{\mathrm{SNIS}}_i f(H(E_i)) \tag{SNIS}
\end{align}
The effectiveness of the importance sampling estimator $\tilde{\mu}$ is theoretically defined as the effective sampling size
$$ESS \triangleq N \frac{Var[\mu(f)]}{Var[\tilde{\mu}(f)]}$$ and often approximated for SNIS estimate $\tilde{\mu}$ as $\frac{1}{\sum_{i=1}^N (w^{\mathrm{SNIS}}_i)^2}$~\cite{kong1992note}. ESS represents the number of i.i.d samples from $\unif$ required to obtain a Monte Carlo estimator $\tilde{\mu}$ with the same accuracy as that of the uniform estimator $\mu$.

\vspace*{-1em}
\section{Performance Evaluation}\label{sec:emp}\vspace*{-1em}
In order to evaluate the effectiveness of \Cref{alg:gen}, we generate multiple random hypergraphs, project the random hypergraphs into simple unweighted graphs, and empirically estimate $(\tilde{\mu}(CC))$ the \textit{average clustering coefficient} ($CC$) on the projected graphs. For simplicity, \textit{we use the alias \textit{SNIS} to imply the algorithmic pipeline of generating several sample hypergraphs using Algorithm~\ref{alg:gen} and then applying the estimator of Equation~\ref{eqn:SNIS}. }

In order to evaluate the efficiency of our generation algorithm (\Cref{alg:gen}), we measure the CPU-time to generate a certain number of random hypergraphs on different datasets. \vspace*{-1em}
\subsection{Datasets} 
We use six graphs and two hypergraph datasets to evaluate the performance of algorithm 2 and compare with that of the MCMC algorithm.

\paragraph{Graphs.} We use the pseudo-fractal family of scale-free simple graphs~\cite{pseudofractal}. Pseudo-fractal graphs are a family $(G_t)$, for integer $t$, of simple graphs where every graph $G_t$ has $3^t, \ldots,3^2,3,3$ vertices of degree $2, \ldots,2^t,2^{t+1}$ respectively. The average clustering coefficient $CC_{t}$ of graph $G_t$ is $\frac{4}{5} \frac{6^t + 3/2}{2^t(2^t+1)}$ and approaches $4/5$ as $t$ grows~\cite{pseudofractal}. 
We are unaware of any analytical form for the average clustering coefficient of projected random graphs\footnote{Parallel-edges are lost after projection} generated following the same degree sequence as $G_t$. However, we observe (\Cref{fig:ccest}) that the empirical expected value of $CC_t$ converges to $\sim0.27$ as $t$ grows.
We construct six graphs $\{G_1,\ldots,G_6\}$ from this family. $\{G_1,\ldots,G_6\}$ have degree sequences of sizes $6$, $15$, $42$, $123$, $366$ and $1095$ respectively, and dimension sequence of sizes $9$,$27$,$81$,$243$,$729$ and $2187$ respectively. The dimension sequence of each graph is a sequence of 2's.
 
\paragraph{Hypergraphs.}
We use the Enron email correspondences and legislative bills in US congress as hypergraph datasets~\cite{benson2018simplicial}\footnote{\url{https://www.cs.cornell.edu/~arb/data/}}.  In \textit{Enron} dataset, the vertices are email addresses at Enron and an edge is comprised of the sender and all recipients of the email. The degree and dimension sequences are of sizes $4423$ and $15653$ respectively. In \textit{congress-bills} dataset, the vertices are congresspersons and an edge is  comprised of the sponsor and co-sponsors (supporters) of a legislative bill put forth in both the House of Representatives and the Senate. The degree and dimension sequences are of sizes $1718$ and $260851$ respectively.
\subsection{Competing Algorithms}
We compare the performance of \textit{SNIS algorithm}, i.e. the SNIS estimator built on our random generation algorithm, with the \textit{MCMC algorithm}~\cite{chodrow2019configuration}. We make two design choices regarding the MCMC algorithm. At first, as the choice for initial hypergraph, we use our construction algorithm~\ref{alg:gen}. Secondly, as the choice for how many iterations to run the Markov chain, we perform autocorrelation analysis on the Markov chain to select a lag value $l$. After selecting $l$, we select random hypergraphs from the chain at every $l$-th hop until required number of hypergraphs are generated. Following standard autocorrelation analysis on Markov chain literature~\cite{cowles1996markov}, $l$ is selected as the lag at which  the autocorrelation function of average clustering coefficient estimate drops below $0.001$. On datasets $G_1$,$G_2$,$G_3$,$G_4$,$G_5$,$G_6$, Enron and congress-bills, we observed and used lag values of $17$, $23$, $115$, $129$, $90$, $304$, $9958$ and $905$ respectively.

\subsection{Effectiveness}
\begin{figure}[t!]
    \centering
    \includegraphics[width=0.8\textwidth]{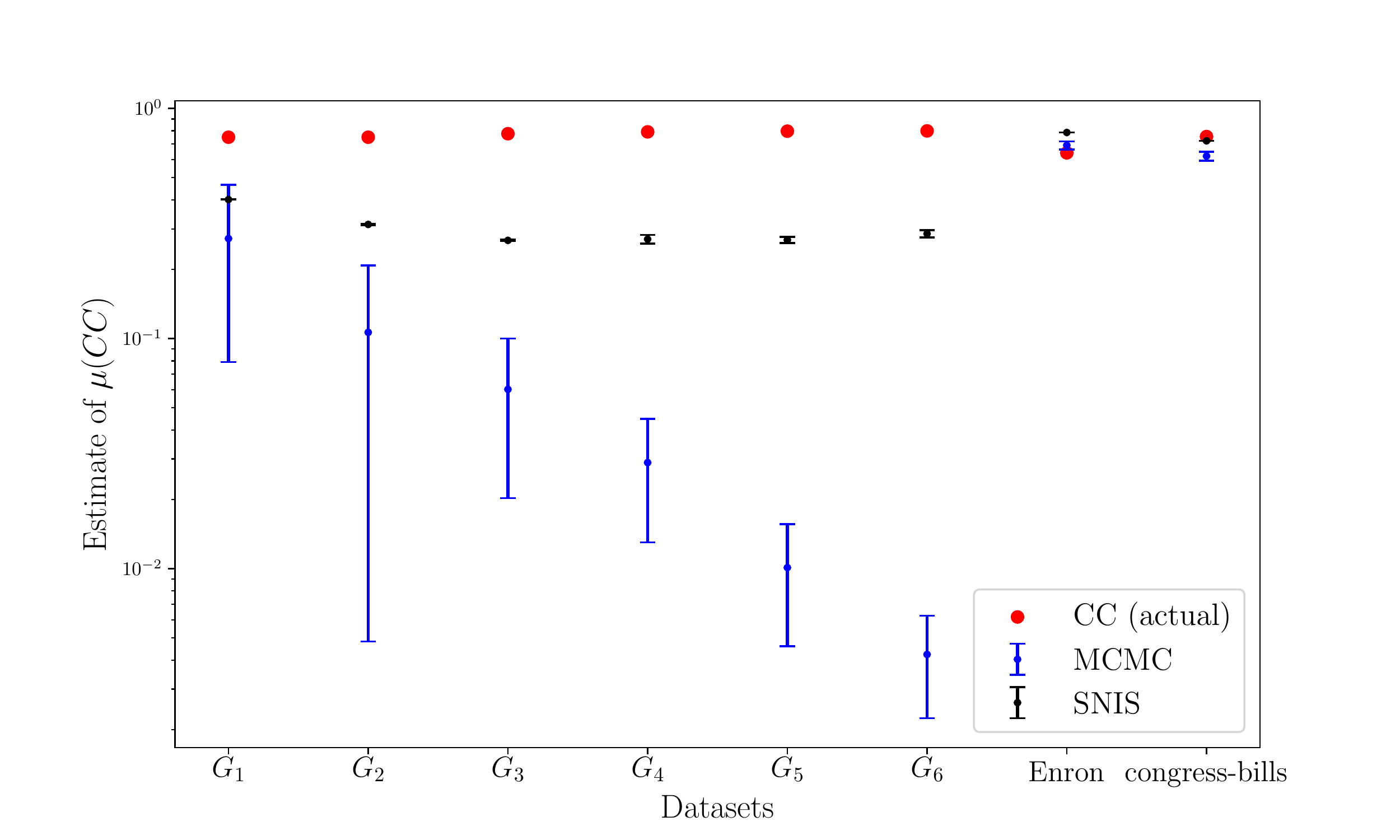}
    \caption{Average clustering coefficients (in log-scale) of the projected random hypergraphs of different datasets and corresponding estimates $\mu(CC)$ using SNIS and MCMC algorithms.}
    \label{fig:ccest}\vspace*{-1em}
\end{figure}

\paragraph{Comparative analysis of estimates of $\mu(CC)$.}
On graph dataset $G_1$ - $G_6$, we construct $500$ random graphs (without loops) using both MCMC and Algorithm~\ref{alg:gen}. On dataset \textit{Enron} and \textit{congress-bills}, we generate $100$ and $20$ random hypergraphs (without loops) respectively using both MCMC and Algorithm~\ref{alg:gen}.  In Figure~\ref{fig:ccest}, we illustrate $CC$ estimates derived using SNIS, MCMC and the actual dataset.  
In Figure~\ref{fig:ccest}, we observe that on average after projection the value of clustering coefficient of the multi-graph is much less than that of a simple graph. We also observe that the average clustering coefficient for the hypergraphs empirically converge to $0.27$ while the average clustering coefficient of corresponding simple graphs converge to $0.8$. This observation is rather expected as parallel-edges decrease the number of triadic closures that would have existed in simple graph. We also observe that, the standard deviation of the SNIS estimates are in significantly smaller than that of the MCMC estimates and closer to the $CC$ of actual data. On Enron and congress-bills hypergraphs, MCMC and SNIS yield comparable estimate for CC.
\textit{Figure~\ref{fig:ccest} indicates that in practice the efficiency and stability of SNIS is either competitive or better than that of MCMC.}

\paragraph{Effective Sample Sizes of estimates of $\mu(CC)$.}
Effective sample size (ESS) (\Cref{sec:sis}) represents the number of i.i.d samples from an uniform sampler required to obtain a uniform Monte Carlo estimator with the same accuracy as that of the SNIS estimator $E_p$. ESS of SNIS algorithm is approximated by $(\sum_{i=1}^N (w^{\mathrm{SNIS}}_i)^2)^{-1}$. The ESS of MCMC samples is defined as $\frac{N}{1+2\sum_{l=1}^\infty \rho(CC^l)}$~\cite{cowles1996markov}, where $\rho(CC^l)$ is the \textit{autocorrelation} function at lag $l$. We consider the summation up-to the lag value for which the autocorrelation drops less than $0.001$. 
We compute the ESS of estimate of $CC$ from both MCMC and SNIS algorithms and and plot them in \Cref{fig:ess}. 

In \Cref{fig:ess}, we observe that the SNIS estimate of $CC$ exhibits higher effective sample size than the estimate using MCMC algorithm. \textit{This observation implies that one can estimate $CC$ using much less number of SNIS samples than MCMC samples.} Although the distinction is not much when the hypergraphs are dense, as apparent from similar values of SNIS for graphs $G_4$,$G_5$ and $G_6$.

\subsection{Efficiency} 
\begin{figure}[t!] \hspace*{-1em}   
 \begin{minipage}{0.5\textwidth}
    \hspace{-2.2em}
    \includegraphics[width=1.2\textwidth]{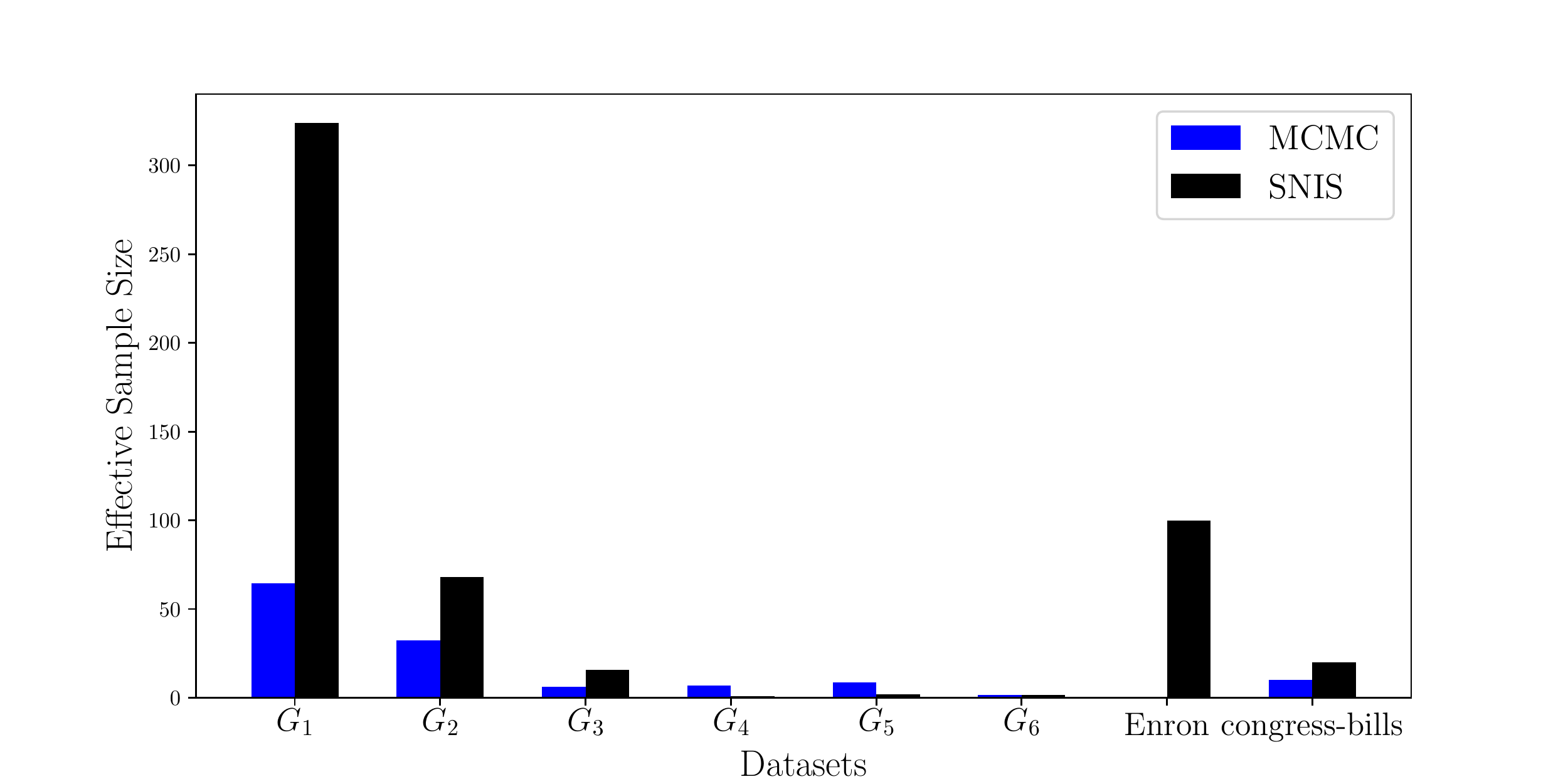}
    \caption{Effective sample sizes of SNIS and MCMC algorithms on $G_1$-$G_6$, Enron and congress-bills datasets. Higher effective sample size indicates better quality of samples.}
    \label{fig:ess}
 \end{minipage}
 \hspace*{1em}
 \begin{minipage}{0.5\textwidth}
     \hspace{-1.5em}
    \includegraphics[width=1.2\textwidth]{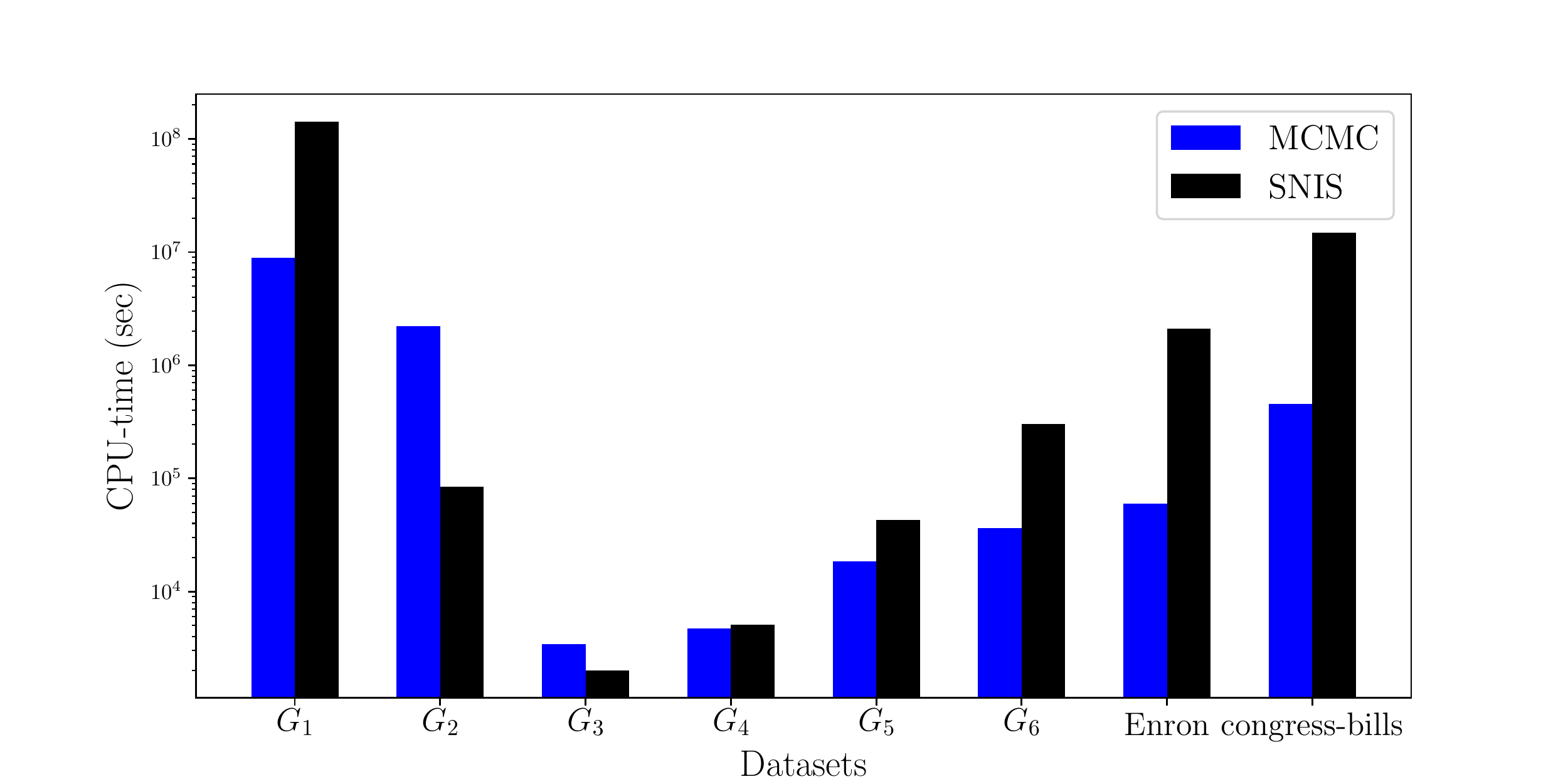}
    \caption{CPU-time (in second, log-scale) to generate $500$ hypergraphs for $G_1$-$G_6$, $100$ hypergraphs for Enron and $20$ hypergraphs for congress-bills datasets. Lower CPU-time is better.}
    \label{fig:tims}
 \end{minipage}\vspace*{-1em}
\end{figure}

We measure the total CPU time (in seconds) taken by the MCMC and ~\Cref{alg:gen} to generate 500 random graphs for the datasets $G_1$-$G_6$, 100 random hypergraphs for Enron dataset, and 20 random hypergraphs for congress-bills datasets respectively. We plot the CPU times in \Cref{fig:tims} for the datasets under consideration.

In \Cref{fig:tims}, we observe that the MCMC algorithm is time-efficient than \Cref{alg:gen}. In particular, it takes less CPU time in generating random hypergraphs with relatively large number of vertices and edges. However, since each run of ~\Cref{alg:gen} generates hypergraphs independently from previous runs, several such hypergraphs can be generated in parallel for the purpose of property estimation. However, such generation is not possible using MCMC algorithm, as previously generated hypergraph are used to switch edges and generate a new hypergraph. We leave potential parallelism as a future work.

\section{Conclusion}\label{sec:conc}
We present two algorithms for construction as well as random generation of hypergraphs with prescribed degree and dimension sequences.

Our algorithms leverage the transposition of properties and algorithms devised for $(0,1)$-matrices with prescribed row- and column-sums to hypergraphs. 
We prove the correctness of the proposed algorithms. We also prove that the generation algorithm generates any random hypergraph following prescribed degree and dimension sequences with non-zero probability.

We propose a self-normalised importance sampling (SNIS) estimator to estimate hypergraph properties and use it to empirically evaluate the effectiveness of random generation.

We compare the effectiveness of the generation algorithm by comparing the SNIS  and MCMC estimates of the average clustering coefficient of the projected graphs obtained from the family of hypergraphs having prescribed degree and dimension sequences. As another measure of quality, we compare the effective sample sizes of the SNIS and MCMC estimates. 

Experimental results reveal that the SNIS estimates are often more accurate and stable at estimating the average clustering coefficient and have higher effective sample sizes compared to the MCMC estimates. Although the present implementation of our generation algorithm takes longer to generate the same number of samples than the MCMC algorithm, we are currently devising a parallel version of our algorithm.

    
\bibliography{biblio} 
\bibliographystyle{alpha}
\newpage
\appendix
\begin{center}
    \Large{\textbf{Supplementary Material}}
\end{center}
\section{Correctness of Algorithm~\ref{alg:constr}: Proof of \Cref{thm:cor_cons}}\label{app:a}
\begin{proof}
\noindent\textbf{Base case:} If $m = 0$, the algorithm terminates with a hypergraph with empty edges ($E=\phi$), which is the only hypergraph with $0$ edges and $n$ vertices.

\noindent\textbf{Inductive case:} Suppose $m>0$. 

By inductive hypothesis, the sequences $(a)^j,(b)^j$ are realisable by a hypergraph $H$ with $m$ edges. Take any incidence matrix of $H$. Sort the columns of the incidence matrix such that the sum of 1's in the $1^{st}$ column is $a^j_1$, $2^{nd}$ column is $a^j_2$, and so on until the sum in $n^{th}$ column is $a^j_n$. Sort the rows of the incidence matrix such that the sum of 1's in the $1^{st}$ row is $b^j_1$, $2^{nd}$ row is $b^j_2$, and so on until the sum in $n^{th}$ row is $b^j_n$. Re-ordering rows does not affect the sums across the columns. Denote the resulting matrix by $M$. 


$M$ is a $(0,1)$-matrix with column-sums $(a)_n^j$ and row-sums $(b)_m^j$.
By Gale-Rysers characterisation, the existence of $M$ implies that $(a)_n^j \prec (\bar{b})_n^j$.

The algorithm constructs $(a)^{j+1}$ by reducing the first $b^j_1$ components in $(a)^{j}$ by 1. Denote the indices of those reduced components as $i_1, i_2,\ldots,i_K$, where $K = b^j_1$.

The algorithm constructs $(b)^{j+1}$ by removing the component $b^j_1$ in $(b)^j$. Lemma~\ref{lemma:maj} implies that, the conjugate $(\bar{b})^{j+1}$ of the constructed sequence  $(b)^{j+1}$ is a sequence derived from $(\bar{b})^{j}$ by reducing first $b^j_1$ components by 1. Denote the indices of those reduced components in $(\bar{b})^j$ as $i'_1,i'_2,\ldots,i'_{K'}$. ~\Cref{lemma:maj} implies that $K' = b^j_1 = K$.

$(a)^j,(\bar{b})^j$  are non-increasing sequence with $(a)^j \prec (\bar{b})^j$.  Since $i_1=i'_1,i_2=i'_2,\ldots,i_K=i'_K$, by Fulkerson-Rysers Lemma (\Cref{lemma:gfr}), $(a)^{j+1} \prec (\bar{b})^{j+1}$. 

Since $(a)^{j+1} \prec (\bar{b})^{j+1}$, by Gale-Rysers theorem (\cref{thm:gale}) there is a $(0,1)$-matrix $M'$ with column-sums $(a)^{j+1}$  and row-sums $(b)^{j+1}$. Since $(b)^{j+1}$ has $m-1$ components, the matrix $M'$ has dimension $(m-1) \times n$. 

Take the associated hypergraph $H'$ of $M'$. $H'$ is a hypergraph with $m-1$ edges and $n$ vertices, which completes the proof.
\end{proof}

\section{Correctness of Algorithm~\ref{alg:gen}: Proof of \Cref{thm:cor_gen}}\label{app:b}
\begin{proof}


\noindent\textbf{Base case:} If $m = 0$, the algorithm terminates with a hypergraph with empty edges ($E=\phi$), which is the only hypergraph with $0$ edges and $n$ vertices.


\noindent\textbf{Inductive case:} Suppose $m>0$. By inductive hypothesis sequences $(a)^j_n,(b)^j_m$ are realisable by a hypergraph $H$ with $m$ edges and $n$ vertices. 

Take any incidence matrix of $H$. Sort the columns of the incidence matrix such that the sum of 1's in the $1^{st}$ column is $a^j_1$, $2^{nd}$ column is $a^j_2$, and so on until the sum in $n^{th}$ column is $a^j_n$. Sort the rows of the incidence matrix such that the sum of 1's in the $1^{st}$ row is $b^j_1$, $2^{nd}$ row is $b^j_2$, and so on until the sum in $n^{th}$ row is $b^j_n$. Re-ordering rows does not affect the sums across the columns. Denote the resulting matrix by $M$. 

$M$ is a $(0,1)$-matrix with column-sums $(a)_n^j$ and column-sums $(b)_m^j$. Gale-Rysers characterisation for matrices implies that $(a)^j \prec (\bar{b})^j$. We show that the choices of indices at iteration $j$ by the algorithm are such that after reduction $(a)^{j+1} \prec \bar{b}^{j+1}$.

At iteration $j$, from each of the integer intervals $\{\mathcal{I}^{j}_i\}$ of indices, the algorithm selects $o_1,o_2,\ldots,$ vertices such that after all the critical indices are considered the total vertices selected to construct edge $e_{j}$ is $b_{j}$. Furthermore, it selects at least $n_1,n_2,\ldots$ vertices at or before indices $k_1,k_2,\ldots$ respectively. 

After reducing 1 from the selected indices in $(a)^j$, the resulting sequence $(a)^{j+1}$  must follow the inequality $\sum_{i=1}^k a_i^{j+1} \leq \sum_{i=1}^k \bar{b}_i^{j+1}$ at every index $k \in [1,n-1]$. If $k$ is critical, $\sum_{i=1}^k a_i^{j+1} \leq \sum_{i=1}^k a_i^{j} - n_k = \sum_{i=1}^k \bar{b}^{j+1}$ where the last equality follows from~\Cref{eqn:margin}. If $k$ is not critical $\sum_{i=1}^k a_i^{j+1} < \sum_{i=1}^k a_i^{j} \leq \sum_{i=1}^k \bar{b}^{j+1}$ where the last inequality follows from the law of logical contraposition applied to definition of critical index (\Cref{eqn:critical}). After all the critical indices are considered, $\sum_{i=1}^n a_i^{j+1} = (\sum_{i=1}^n a_i^j) - b^j_1 = (\sum_{i=1}^n b_i^j) - b^j_1 = \sum_{i=1}^n b_i^{j+1}$. The second equality follows from the definition of dominance (Definition~\ref{def:dom}).

By definition of dominance, the inequalities $\sum_{i=1}^k a_i^{j+1} \leq \sum_{i=1}^k \bar{b}_i^{j+1}$ at every index $k \in [1,n-1]$ and the equality $\sum_{i=1}^n a_i^{j+1} = \sum_{i=1}^n b_i^{j+1}$ imply that $(a)^{j+1} \prec (\bar{b})^{j+1}$. Consequently, Gale-Rysers theorem (\Cref{thm:gale}) implies that there exists a $(0,1)$-matrix $M'$ with column-sums $(a)^{j+1}$  and row-sums $(b)^{j+1}$. Since $(b)^{j+1}$ has $m-1$ components, the matrix $M'$ has dimension $(m-1) \times n$. 

Take the associate hypergraph $H'$ of the incidence matrix $M'$. $H'$ has $(m-1)$ edges and $n$ vertices, which completes the proof.

\end{proof}

\section{Non-zero Probability of Randomly Generating any Hypergraph using Algorithm~\ref{alg:gen}: Proof of ~\Cref{thm:prob_gen}}\label{app:c}
\begin{proof}

Let $H_1 = (V,E_1 = \{e_1,\ldots,e_m\})$ be an arbitrary hypergraph realisation of sequences $(a)^1 = (a)_n$ and $(b)^1 = (b)_m$. Without loss of generality, assume the degrees $d^1(v_1) = a_1,d^1(v_2)=a_2,\ldots,d^1(v_n)=a_n$, the dimensions $|e_1|= b_1, |e_2|=b_2,\ldots,|e_m|=b_m$, and $(a)_n$,$(b)_m$ both monotonically non-increasing. Here, $d^1(v_i)$ is the degree of vertex $v_i$ in $H_1$.

Construct hypergraph $H_2 = (V, E_2 = E_1 \setminus \{e_1\}$ by removing edge $e_1$ from $E_1$. Let $e_1 = \{v_{k_1},v_{k_2},\ldots,v_{k_{b_1}}\}$ where $k_1,k_2,\ldots,k_{b_1}$ are the indices of vertices $v_{k_1},v_{k_2},\ldots,v_{k_{b_1}}$ in $(a)^1$. The degree $d^2(v_i)$ of vertex $v_i$ in $H_2$ is the following 

\[
d^2(v_i) = \begin{cases}
d^1(v_i) - 1 \quad \text{ if } v_i \in e_1,\\
d^1(v_i) \quad \text{otherwise.} 
\end{cases}
\]
By construction, $H_2$ has degree sequence $(a)^2 = (d^2(v_1),d^2(v_2),\ldots,d^2(v_n))$ and dimension sequence $(b)^2 = (b_2,b_3,\ldots,b_m)$. Since $H_2$ is a hypergraph realisation of sequnces $(a)^2$ and $(b)^2$, Gale-Rysers theorem implies that the $(a)^2 \prec (\bar{b})^2$.

We show that, the algorithm, at iteration $1$, considers all possible choices for indices $k_1,k_2, \ldots, k_{b_1}$ such that $(a)^2 \prec (\bar{b})^2$ after iteration $1$. In other words the probability $\prob(e_1)$ with which it constructs edge $e_1$ is non-zero.


To show that $\prob(e_1)>0$, we explicitly compute $\prob(e_1)$.
At iteration 1, The algorithm selects $o_1$ vertices from $[1,k^1_1]$ with probability $\frac{o_1}{k^1_1 (\min(b^1_1,k^1_1) - n^1_1 + 1)}$. Conditioned on the fact that $o_1$ vertices are selected, the algorithm subsequently selects $o_2$ vertices from $[k^1_1+1,k^1_2]$ with probability  $\frac{o_2}{(k^1_2-k^1_1)(\min(k^1_2-k^1_1,b^1_1-o_1) - (n^1_2-o_1) + 1)}$ Continue to compute the probabilities associated to every selection $o_i>0$ of vertices. Multiplying all the conditional probabilities, we get the joint probability  $$\prob(e_1) = \frac{o_1}{k^1_1(\min(b^1_1,k^1_1) - n^1_1 + 1)}\frac{o_2}{(k^1_2-k^1_1)(\min(k^1_2-k^1_1,b^1_1-o_1) - (n^1_2-o_1) + 1)}\ldots$$ $\prob(e_1)$ is non-zero as each term in the product is non-zero.



Continue to remove the edges $e_2,e_3,\ldots,e_m$ in that order and compute the probabilities $\prob(e_2|e_1),\ldots,\prob(e_m|e_{m-1},\ldots,e_1)$. Again, all these probabilities are non-zero, by arguments similar to that of $\prob(e_1)$.

Finally, construct $H_{m+1}$, which is the empty hypergraph $(V, E_{m+1} = \phi)$. The algorithm constructs $H_{m+1}$ with probability 1, since it terminates with $E = \phi$ without entering the for-loop when $m=0$. 

\textit{The joint probability with which the algorithm constructs the edge-sequence $E_1 \triangleq \lbrace e_1, \ldots, e_m \rbrace$} is denoted as
$$\prob(E_1) \triangleq \prob(e_1) \prob(e_2|e_1) \ldots  \prob(e_m|e_{m-1},\ldots,e_1).$$
The product is non-zero, as each term in the product is non-zero.

There are $\frac{m!}{\prod_j mult^{E_1}(e_j)!}$ permutation of $e_1,\ldots,e_m$ that results in the same hypergraph as $H_1$. Here, $mult^{E_1}(e_j)$ is the multiplicity of edge $e_j$ in multiset $E_1$. 

Let us denote the set of all permutations of $E_1$ by $[E_1]$. Thus the probability $\prob(H_1)$ at which the algorithm constructs $H_1$ is given by $\sum_{E \in [E_1]} \prob(E)$. $\prob(H_1)$ is non-zero, as each term in the summation is non-zero.

Since $H_1$ was an arbitrary chosen hypergraph, the argument is true for any hypergraph realisation of $(a)_n$ and $(b)_m$.



\end{proof}

\section{Importance Sampling Estimator $\hat{\mu}(f)$ is an Unbiased Estimator}\label{app:d}
In this section, we prove that the importance sampling estimator $\hat{\mu}(f)$ in Equation~\eqref{eq:is} is an unbiased estimator of $\mu(f)$, where $f:\mathcal{H}_{ab} \rightarrow \mathbb{R}$ is a hypergraph property. As \Cref{alg:gen} constructs a hypergraph by constructing a sequence of edges, we distinguish between a hypergraph and a sequence of edges using the following definition. 

\begin{definition}[Equivalence class of an edge-sequence]
For any sequence $E \in \mathcal{E}_{ab}$ of edges, let $H(E)$ be the corresponding hypergraph in $\mathcal{H}_{ab}$. We call $E,E' \in \mathcal{E}_{ab}$ equivalent if $H(E) = H(E')$, and denote the relation by $E \sim E'$. The equivalence class of edge sequence $E \in \mathcal{E}_{ab}$ is the set $[E] \triangleq \{E' \in \mathcal{E}_{ab}: E' \sim E \}$. 
\end{definition}

\begin{definition}[Induced functions on an edge-sequence]\label{def:edgeseq}
Given a function $f: \mathcal{H}_{ab} \rightarrow \mathbb{R}$, the induced function $\hat{f}$ on the larger space $\mathcal{E}_{ab}$ is defined as $\hat{f}(E) \triangleq f(H(E)) = f(H)$. The probability mass function $\hat{\mathcal{U}}$ on $\mathcal{E}_{ab}$ induced by the uniform distribution $\mathcal{U}:\mathcal{H}_{ab} \rightarrow [0,1]$ is defined as $\hat{\mathcal{U}}(E) \triangleq \frac{\mathcal{U}(H(E))}{|[E]|}$.
\end{definition}

We state and prove the following theorem which says that, despite estimating hypergraph property $f$ by sampling edge-sequences according to $\prob$, rather than $\mathcal{U}$, the importance sampling estimator $\hat{\mu}$ is unbiased. 

\begin{theorem}
Let $H \in \mathcal{H}_{ab}$ be a random hypergraph drawn according to $\mathcal{U}$. Let $E \in \mathcal{E}_{ab}$ be the sequence of edges drawn according to $\prob$ by \Cref{alg:gen}. Then 

$$ \mathbb{E}_\mathcal{U}[f(H)] = 
 \mathbb{E}_{\prob} \left[ \frac{\hat{\mathcal{U}}(E)}{\prob(E)} \hat{f}(E) \right] 
$$
\end{theorem}
and in particular, for $E_1,\ldots,E_N$ the output sequences of N independent runs of \Cref{alg:gen} 

$$
\hat{\mu} = \frac{1}{N} \sum_{i=1}^{N} \frac{\hat{\mathcal{U}}(E_i)}{\prob(E_i)} \hat{f}(E_i)
$$ is an unbiased estimator of $\mu \triangleq \mathbb{E}_\mathcal{U}[f(H)]$

\begin{proof} 
\begin{align}
\mathbb{E}_\mathcal{U}[f(H)] &= \sum_{H \in \mathcal{H}_{ab}} f(H)\mathcal{U}(H) \notag \\ 
 &=  \sum_{H \in \mathcal{H}_{ab}} \mathcal{U}(H) \sum_{E' \in [E]}  \frac{f(H(E'))}{|[E]|}  \label{eqn:line2}\\
 &= \sum_{H \in \mathcal{H}_{ab}} \mathcal{U}(H)\sum_{E' \in [E]}  \frac{\hat{f}(E')}{|[E]|}  \label{eqn:line3}\\
&= \sum_{H \in \mathcal{H}_{ab}} \sum_{E' \in [E]} \frac{\hat{f}(E')\mathcal{U}(H)}{|[E]|}  \notag \\
&= \sum_{H \in \mathcal{H}_{ab}} \sum_{E' \in [E]} \frac{\hat{f}(E')\mathcal{U}(H(E'))}{|[E]|}  \notag \\
&= \sum_{H \in \mathcal{H}_{ab}} \sum_{E' \in [E]} \frac{\hat{f}(E')(\hat{\mathcal{U}}(E') |[E']|)}{|[E]|}  \label{eqn:line5}\\
&= \sum_{E \in \mathcal{E}_{ab}}  \hat{f}(E)\hat{\mathcal{U}}(E) \label{eqn:line6} \\
&= \sum_{E \in \mathcal{E}_{ab}}  \frac{\hat{f}(E)\hat{\mathcal{U}}(E)}{\prob(E)} \prob(E) \notag\\
&=  \mathbb{E}_{\prob} \left[ \frac{\hat{\mathcal{U}}(E)}{\prob(E)} \hat{f}(E) \right] \notag
\end{align}
Equation~\eqref{eqn:line2} is obtained due to the fact that, $f(H(E')) = f(H)$ for any $E' \sim E$. Equation~\eqref{eqn:line3} follows from the definition of $\hat{f}$ (Definition~\ref{def:edgeseq}). Equation~\eqref{eqn:line5} follows from the definition of $\hat{\mathcal{U}}$ (Definition~\ref{def:edgeseq}). Equation~\eqref{eqn:line6} follows from Equation~\eqref{eqn:line5} as $[E'] = [E]$ for any $E' \sim E$ and the terms inside the double summation do not involve $H$. 

\end{proof}
\end{document}